\documentclass[11pt,twoside]{article}
\usepackage{amsfonts}
\usepackage{amssymb}
\usepackage{graphicx,color}
\usepackage{a4wide}
\usepackage{amsmath}
\usepackage{amsthm}

\graphicspath{{./Figs/}}
\DeclareGraphicsExtensions{.pdf,.jpg,.png,.eps,.gif}
\date{\today}

\newtheorem{theorem}{Theorem}[section]
\newtheorem{lemma}[theorem]{Lemma}

\newtheorem{conjecture}[theorem]{Conjecture}

\theoremstyle{definition}
\newtheorem{definition}[theorem]{Definition}
\newtheorem{example}[theorem]{Example}
\newtheorem{remark}[theorem]{Remark}

\begin{document}

\vspace*{5mm}

\noindent
\textbf{\LARGE Linear Programming Bounds}

\date{}

\vspace*{5mm}
\noindent
\textsc{Peter Boyvalenkov} \hfill \texttt{peter@math.bas.bg} \\
{\small Institute for Mathematics and Informatics, Bulgarian Academy of Sciences, Sofia, Bulgaria;} \\
{\small Technical Faculty, Southwestern University, Blagoevgrad, Bulgaria} \\[3pt]
\textsc{Danyo Danev} \hfill \texttt{danyo.danev@liu.se} \\
{\small Department of Electrical Engineering and Department of Mathematics, \\
	Link\"oping University, SE-581 83 Link\"oping, Sweden} 

\medskip

\begin{abstract}
This chapter is written for the forthcoming book "A Concise Encyclopedia of Coding Theory" (CRC press), edited by W. Cary Huffman, Jon-Lark Kim, and Patrick Sol\'e. This book will collect short but foundational articles, emphasizing definitions, examples, exhaustive references, and basic facts. The target audience of the Encyclopedia is upper level undergraduates and graduate students. 
\end{abstract}

\section{Preliminaries -- Krawtchouk polynomials, codes, and designs}
\label{LPB:s1}

\begin{definition}[Krawtchouk polynomials]
	\index{Krawtchouck polynomial $K_k^{(n,q)}(x)$}
	\label{kr-1}
	{\rm Let $n \geq 1$ and $q \geq 2$ be integers. The Krawtchouk polynomials 
	are 
	defined as
	\[   		 K_i^{(n,q)}(z) :=
		 \sum_{j=0}^i (-1)^{j}(q-1)^{i-j}q^{j} \binom{n-j}{n-i}\binom{z}{j},
		 \ \ i=0,1,2,\ldots ,	\]
	where $\binom{z}{j} := z(z-1)\cdots(z-j+1)/j!$, $z \in \mathbb{R}$.}
\end{definition}

\begin{theorem}[\cite{lev95}]
	\index{Krawtchouck polynomial $K_k^{(n,q)}(x)$!three-term reccurence}
	\label{kr1-t1}
	The polynomials $K_i^{(n,q)}(z)$ satisfy the three-term recurrence relation
	\[  		(i+1)K_{i+1}^{(n,q)}(z) = [i+(q-1)(n-i)-qz]K_i^{(n,q)}(z) -
		 (q-1)(n-i+1)K_{i-1}^{(n,q)}(z) 	\]
	with initial conditions $K_0^{(n,q)}(z)=1$ and $K_1^{(n,q)}(z)=n(q-1)-qz$.
\end{theorem}

\begin{theorem}[\cite{lev95}]
	\index{inner product!Hamming spaces}
	\index{discrete measure}
	\label{kr1-t2}
	The discrete measure
	\begin{equation}
		\label{KrawOrtho}
		d\mu_n (t) := q^{-n}\sum_{i=0}^n \binom{n}{i} (q-1)^i \delta(t-i) dt,
	\end{equation}
	where $\delta(i)$ is the Dirac-delta measure at $i \in \{0,1,\ldots,n\}$, 
	and the form
	\begin{equation}
		\label{InnerProd}
		\langle f,g \rangle=\int f(t) g(t) d\mu_n (t)
	\end{equation}
	define an inner product over the class $\mathcal{P}_n$ of real polynomials of 
	degree less than or equal to $n$.
\end{theorem}

\begin{theorem}[Orthogonality relations, \cite{lev95}]
	\index{polynomial!orthogonality}
	\label{or1}
	Under the inner product \eqref{InnerProd} the Krawtchouk polynomials satisfy
	\[ 
		\sum_{u=0}^n K_i^{(n,q)}(u) K_j^{(n,q)}(u)(q-1)^u \binom{n}{u} = 
		\delta_{i,j} q^n(q-1)^i \binom{n}{i} 
	\]
	for any $i,j \in \{0,1,\ldots,n\}$. Moreover, 
	\[ 
		\sum_{u=0}^n K_i^{(n,q)}(u) K_u^{(n,q)}(j)=\delta_{i,j} q^n. 
	\]
\end{theorem}

\begin{theorem} [Expansion, \cite{lev95}]
	\index{polynomial!expansion}
	\label{exp-1}
	If $f(z)=\sum_{i=0}^n f_i K_i^{(n,q)}(z)$, then 
	\begin{eqnarray*}
		f_i &=& \left[ q^n(q-1)^i \binom{n}{i} \right]^{-1}  \sum_{u=0}^n f(u) 
				K_i^{(n,q)}(u)(q-1)^u \binom{n}{u} \\
			&=& q^{-n} \sum_{u=0}^n f(u)K_u^{(n,q)}(i).
	\end{eqnarray*}
\end{theorem}

\begin{definition}
	{\rm Define $K_{n+1}^{(n,q)}(z):=\frac{q^{n+1}}{(n+1)!} \prod_{u=0}^n 
	(u-z)$. Note that $K_{n+1}^{(n,q)}(z)$ is orthogonal to any polynomial 
	$K_i^{(n,q)}(z)$, $i=0,1,\ldots,n$, with respect to the measure 
	\eqref{KrawOrtho}.}
\end{definition}

\begin{theorem} [Krein condition, \cite{lev95}]
	\index{Krein condition}
	\label{krein}
	For any $i,j \in \{0,1,\ldots,n\}$ 
	\[ 
		K_i^{(n,q)}(z)K_j^{(n,q)}(z) = \sum_{u=0}^n p_{i,j}^{u} K_u^{(n,q)}(z) 
		\pmod{K_{n+1}^{(n,q)}(z)} 
	\]
	with $p_{i,j}^u=0$ if $i+j>u $, $p_{i,j}^u>0$ if $i+j=u \leq n$, and 
	$p_{i,j}^u \geq 0$ otherwise. 
\end{theorem}

\begin{definition}
	\label{Tk-kernels}
	{\rm Denote 
	\[ 
		T_i^{n,q}(z,w) := \sum_{j=0}^i K_j^{(n,q)}(z) K_j^{(n,q)}(w)   
		\left( (q-1)^j \binom{n}{j} \right)^{-1}. 
	\]}
\end{definition}

\begin{theorem}[Christoffel-Darboux formula, \cite{lev95}]
	\index{formula!Christoffel-Darboux}
	For any \\ $i=0,1,\ldots,n$ and any real $z$ and $w$ 
	\[ 
		(w-z)T_i^{n,q}(z,w) = \frac{i+1}{q(q-1)^i \binom{n}{i}} 
		\left(K_{i+1}^{(n,q)}(z)K_{i}^{(n,q)}(w) - 
		K_{i+1}^{(n,q)}(w)K_{i}^{(n,q)}(z)\right). 
	\]
\end{theorem}

Let $\mathcal{C} \subseteq F_q^n$ be a code, where $F_q=\{0,1,\ldots,q-1\}$ is 
the alphabet of $q$ symbols (so $q$ is not necessarily a power of a prime). For 
$x,y \in F_q^n$, recall that $d(x,y)$ is the number of coordinates where $x$ 
and $y$ disagree. 

\begin{definition}
	\index{code!distance distribution}
	{\rm The vector $B(\mathcal{C})=(B_0,B_1,\ldots,B_n)$, where 
	\[ 
		B_i=\frac{1}{|\mathcal{C}|} \left| \{ (x,y) \in \mathcal{C}^2 : 
		d(x,y)=i\}\right|, \ \  i=0,1,\ldots,n ,
	\]
	is called the {\bf distance distribution} of $\mathcal{C}$. Clearly, 
	$B_0=1$ and $B_i=0$ for $i=1,2,\ldots,d-1$, where $d$ is the minimum 
	distance of $\mathcal{C}$.}
\end{definition}

\begin{definition}
	\index{code!dual distance distribution}
	\index{MacWilliams transform}
	{\rm The vector $B^\prime(\mathcal{C}) = 
	(B^\prime_0,B^\prime_1,\ldots,B^\prime_n)$, where 
	\[ 
		B^\prime_i=\frac{1}{|\mathcal{C}|} \sum_{j=0}^n B_j K_i^{(n,q)}(j), \ \ 
		i=0,1,\ldots,n ,
	\]
	is called the {\bf dual distance distribution} of $\mathcal{C}$ or the {\bf 
	MacWilliams transform} of $B(\mathcal{C})$. Obviously $B^\prime_0=1$.  }
\end{definition}

\begin{theorem}[\cite{del73b,del73a}]
	The dual distance distribution of $\mathcal{C}$ satisfies
	\[ 
		B^\prime_i \geq 0, \ \ i=1,2,\ldots,n. 
	\] 
\end{theorem}

\begin{theorem}[\cite{macslo77}]
	If $q$ is a power of a prime and $\mathcal{C}$ is a linear code in 
	$\mathbb{F}_q^n$, then $B^\prime(\mathcal{C})$ is the distance distribution 
	of the dual code $\mathcal{C}^\perp$.  
\end{theorem}

\begin{definition}
	\index{distance!dual}
	\index{distance!external}
	\label{ddist}
	{\rm The smallest positive integer $i$ such that $B^\prime_i \neq 0$ is 
	called the {\bf dual distance} of $\mathcal{C}$ and is denoted by 
	$d^\prime=d^\prime(\mathcal{C})$. Denote by $s=s(\mathcal{C})$ (resp., 
	$s^\prime=s^\prime(\mathcal{C})$) the number of nonzero $B_i$'s (resp., 
	$B^\prime_i$'s), $i \in \{1,2,\ldots,n\}$, i.e. 
	\[ 
		s=|\{ i : B_i \neq 0, i>0\}|,  
		\ s^\prime=|\{ i : B^\prime_i \neq 0, i>0\}|. 
	\] 
	The number $s^\prime$ is called the {\bf external distance} of 
	$\mathcal{C}$. Define $\delta=0$ if $B_n=0$ and $\delta=1$ otherwise 
	(respectively, $\delta^\prime=0$ if $B^\prime_n=0$ and $\delta^\prime=1$ 
	otherwise).}
\end{definition}

\begin{definition}
	\index{code!$\tau$-design}
	\index{code!strength}
	\index{code!index}
	{\rm Let $\mathcal{C} \subseteq F_q^n$ be a code and $M$ be a codeword 
	matrix consisting of all vectors of $\mathcal{C}$ as rows. Then 
	$\mathcal{C}$ is called a $\tau${\bf -design} if any set of $\tau$-columns 
	of $M$ contains any $\tau$-tuple of $F_q^\tau$ the same number of times 
	(namely, $\lambda:=|\mathcal{C}|/q^\tau$). The largest positive integer 
	$\tau$ such that $\mathcal{C}$ is a $\tau$-design is called the
	{\bf strength} of $\mathcal{C}$ and is denoted by $\tau(\mathcal{C})$. The 
	number $\lambda$ is called the {\bf index} of $\mathcal{C}$. }
\end{definition}

\begin{remark}
	\index{orthogonal array}
	{\rm A $\tau$-design in $F_q^n$ is also called an {\bf orthogonal array of 
	strength} $\tau$ \cite{del73a} or a $\tau${\bf{}-wise independent set} 
	\cite{alogolhasper92} (see also \cite{hedslostu99}).  }
\end{remark}

\begin{theorem}[\cite{del73b,del73a}]
	If $\mathcal{C} \subseteq F_q^n$ has dual distance $d^\prime = 
	d^\prime(\mathcal{C})$, then $\mathcal{C}$ is a $\tau$-design, where 
	$\tau=d^\prime-1$.
\end{theorem}

\begin{definition}
	\index{polynomial!dual}
	\label{poly-duality}
	{\rm For a real polynomial $f(z)=\sum_{i=0}^n f_i K_i^{(n,q)}(z)$, the 
	polynomial
	\[ 
		\widehat{f}(z)=q^{-n/2} \sum_{j=0}^n f(j)K_j^{(n,q)}(z) 
	\]
	is called the {\bf dual} to $f(z)$. Note that the dual to $\widehat{f}(z)$ 
	is $f(z)$ and also that $\widehat{f}(i)=q^{n/2}f_i$ for any 
	$i\in\{0,1,\dots,n\}$.}
\end{definition}

\section{General linear programming theorems}
\label{LPB:s2}

\begin{theorem}[\cite{del73b,del73a}]
	\label{lpb-q-codes}
	For any code $\mathcal{C} \subseteq F_q^n$ with distance distribution 
	$(B_0,B_1,\ldots,B_n)$ and  dual distance distribution 
	$(B^\prime_0,B^\prime_1,\ldots,B^\prime_n)$, and any real polynomial 
	$f(z)=\sum_{i=0}^n f_i K_i^{(n,q)}(z)$, it is valid that
	\[ 
		f(0) + \sum_{i=1}^n B_i f(i) = |\mathcal{C}|
		\left( f_0+\sum_{i=1}^n f_i B^\prime_i \right). 
	\]
\end{theorem}

In Definition 1.9.1 in Chapter 1, $A_q(n,d)$ is defined for 
codes over $\mathbb{F}_q$. We now extend that definition to codes over $F_q$ as 
only the alphabet size is important and not its structure.

\begin{definition}
	\index{code!optimal!$A_q(n,d)$}
	{\rm For fixed $q$, $n$, and $d \in \{1,2,\ldots,n\}$ denote 
	\[ 
		A_q(n,d) := \max \{ |\mathcal{C}|: \mathcal{C} \subseteq F_q^n, \ 
		d(\mathcal{C})=d \}.   
	\]}
\end{definition}

\begin{definition}
	\index{code!optimal!$B_q(n,\tau)$}
	{\rm For fixed $q$, $n$, and $\tau \in \{1,2,\ldots,n\}$ denote 
	\[ 
		B_q(n,\tau) := \min \{ |\mathcal{C}|: \mathcal{C} \subseteq F_q^n, \ 
		\tau(\mathcal{C})=\tau \}.   
	\]}
\end{definition}

\begin{theorem}[Linear Programming Bound for codes \cite{del73b,del73a}]
	\index{bound!linear programming}
	\label{lp-codes}
	Let the real polynomial $f(z)=\sum_{i=0}^n f_i K_i^{(n,q)}(z)$ 
	satisfy 
	the 
	conditions
	\begin{itemize}
		\item[(A1)] $f_0>0$, $f_i \geq 0$ for $i=1,2,\ldots,n$;
		\item[(A2)] $f(0)>0$, $f(i) \leq 0$ for $i=d,d+1,\ldots,n$.
	\end{itemize}
	Then $A_q(n,d) \leq f(0)/f_0$. Equality holds for codes $\mathcal{C} 
	\subseteq F_q^n$ with $d(\mathcal{C})=d$, distance distribution  
	$(B_0,B_1,\ldots,B_n)$, dual distance distribution 
	$(B^\prime_0,B^\prime_1,\ldots,B^\prime_n)$ and polynomials $f(z)$ such that 
	$B_if(i)=0$ and $B^\prime_if_i=0$ for every $i=1,2,\ldots,n$.
\end{theorem}

\begin{remark}
	{\rm The polynomial $f$ from Theorem \ref{lp-codes} is implicit in Theorem 
	1.9.23 a) from Chapter 1 as $\sum_{i=0}^n 
	B_i K_i^{(n,q)}(i)$ and it is normalized for $f_0$ (or $B_0$) to be equal 
	to 1; note also the normalization of the Krawtchouk polynomials. Thus the 
	bound $f(1)/f_0$ appears as $\sum_{i=0}^n B_i$ in Chapter 1.}
\end{remark}

\begin{theorem} [Linear Programming Bound for designs \cite{del73b,del73a}]
	\index{bound!linear programming}
	\label{lp-designs}
	Let the real polynomial $f(z)=\sum_{i=0}^n f_i K_i^{(n,q)}(z)$ satisfy the 
	conditions
	\begin{itemize}
		\item[(B1)] $f_0>0$, $f_i \leq 0$ for $i=\tau+1,\tau+2,\ldots,n$;	
		\item[(B2)] $f(0)>0$, $f(i) \geq 0$ for $i=1,2,\ldots,n$.
	\end{itemize}
	Then $B_q(n,\tau) \geq f(0)/f_0$. Equality holds for designs $\mathcal{C} 
	\subseteq F_q^n$ with $\tau(\mathcal{C})=\tau$, distance distribution  
	$(B_0,B_1,\ldots,B_n)$, dual distance distribution 
	$(B^\prime_0,B^\prime_1,\ldots,B^\prime_n)$ and polynomial $f(z)$ such that 
	$B_if(i)=0$ and $B^\prime_if_i=0$ for every $i=1,2,\ldots,n$.
\end{theorem}

\begin{remark}
	{\rm The Rao Bound in Theorem 	\ref{r-h-bound}
 and Levenshtein Bound in \ref{l-q-bound}
 below are obtained by suitable polynomials in Theorems 
	\ref{lp-codes} and \ref{lp-designs}, respectively. Examples of codes 
	attaining these bounds are listed in Table \ref{Table1-lev-rao} below.}
\end{remark}

\begin{theorem}[Duality, \cite{lev99}]
	\index{theorem!duality}
	\label{duality-thm}
	A real polynomial $f(z)$ satisfies the conditions (A1) and (A2) if and only 
	if its dual polynomial $\widehat{f}$ satisfies the conditions (B1) and 
	(B2). Moreover,
	\[ 
		\frac{f(0)}{f_0} \cdot \frac{\widehat{f}(0)}{\widehat{f}_0}=q^n . 
	\]
\end{theorem}

\begin{remark}
	\label{duality-rem}
	{\rm Rephrased, the duality means that for any polynomial $f$ which is good 
	for linear programming for codes, its dual is good for linear programming 
	for designs (and conversely). Thus we obtain, in a sense, bounds for free. 
	In particular, the duality justifies the pairs of bounds in Theorems \ref{s-bound}, 
	\ref{r-h-bound} and \ref{l-q-bound} below.}
\end{remark}

See also \cite{brocohneu89,dellev98,god88,god93,lev99}.

\begin{definition}
	\index{code!potential energy} 
	{\rm   For a code $\mathcal{C} \subseteq F_q^n$ and a function $h : 
	\{1,2,\ldots,n\} \to \mathbb{R}$ the {\bf potential energy of $\mathcal{C}$ 
	with respect to} $h$ is defined to be

\[ 
		E_h(\mathcal{C}) := \sum_{x,y \in \mathcal{C}, x \neq y } h(d(x,y)). 
	\]}
\end{definition}

\begin{definition}
	\index{code!optimal!$E_h(n,M;q)$} 
	{\rm For fixed $q$, $n$, $h$, and $M \in \{2,3,\ldots,q^n\}$ denote 
	\[  
		E_h(n,M;q) := \min \{E_h(\mathcal{C}): \mathcal{C} \subseteq F_q^n, \ 
		|\mathcal{C}|=M\}. 
	\]}
\end{definition}

\begin{theorem}[Linear Programming Bound for energy of codes \cite{cohzha14}]
	\index{bound!linear programming}
	\label{LP-hamming-energy}
	Let $n$ and $q$ be fixed, $h:(0,n] \to (0,+\infty)$ be a function, and $M 
	\in \{2,3,\ldots,q^n\}$. Let the real polynomial $f(z)=\sum_{i=0}^n f_i 
	K_i^{(n,q)}(z)$ satisfy the conditions
	\begin{itemize}
		\item[(D1)] $f_0>0$, $f_i \geq 0$ for $i=1,2,\ldots,n$;	
		\item[(D2)] $f(0)>0$, $f(i) \leq h(i)$ for $i=1,2,\ldots,n$.
	\end{itemize}
	Then $E_h(n,M;q) \geq M(f_0M-f(0))$. Equality holds for codes $\mathcal{C} 
	\subseteq F_q^n$ with distance distribution  $(B_0,B_1,\ldots,B_n)$, dual 
	distance distribution $(B^\prime_0,B^\prime_1,\ldots,B^\prime_n)$ and 
	polynomials $f(z)$, such that $B_i\left[f(i)-h(i)\right]=0$ and 
	$B^\prime_if_i=0$ for every $i=1,2,\ldots,n$.
\end{theorem}

\begin{remark}
	{\rm  Theorems \ref{lp-codes}, \ref{lp-designs} and \ref{LP-hamming-energy} 
	can be applied with the usual simplex method for quite large parameters. 
	For instance, the website \cite{sage} (Delsarte, a.k.a. Linear Programming 
	(LP), upper bounds) offers a tool for computation of bounds via integer LP 
	with Theorem \ref{lp-codes}. Several websites maintain tables of best known 
	bounds (lower and upper) for codes of relatively small lengths (see, e.g. \cite{bro18}).}
\end{remark}

\section{Universal bounds}
\label{LPB:s3}

The Singleton Bound presented in Theorem 1.9.10 in Chapter 1 is an upper 
bound on the code cardinality, given $q$, $n$, and $d$. It is the upper bound 
in \eqref{s-bound-pair} below. Its proof by linear programming and the duality 
imply the lower bound in \eqref{s-bound-pair}. 

\begin{theorem}[Singleton Bound \cite{sin64}]
	\index{bound!Singleton}
	\label{s-bound}
	For any code $\mathcal{C} \subseteq F_q^n$ with minimum distance $d$ and 
	dual distance $d^\prime$
	\begin{equation}
		\label{s-bound-pair}
		q^{d^\prime-1} \leq |\mathcal{C}| \leq q^{n-d+1}. 
	\end{equation}
\end{theorem}

The bounds \eqref{s-bound-pair} can be attained only simultaneously and this 
happens if and only if $d+d^\prime=n+2$ and all possible distances are realized 
(so the attaining code is an MDS code). 

\begin{definition} 
	{\rm   For fixed $q$, $n$, and $d$ denote by
	\[  
		V_k(n,q) := \sum_{i=0}^k \binom{n}{i} (q-1)^i, \ 0 \leq k \leq n ,
	\]
	(volume of a sphere of radius $k$ in $F_q^n$) and
	\[ 
		H^{n,q}(d) = q^{\varepsilon}V_k(n-\varepsilon,q), 
	\]
	where $d=2k+1+\varepsilon$, $\varepsilon \in \{0,1\}$.    }
\end{definition}

The Sphere Packing or Hamming Bound presented in Theorem 1.9.6 in Chapter 1 is another upper bound on the code size, given 
$q$, $n$, and $d$. Rao gave a lower bound on the code size, given $q$, $n$, and 
$d^\prime$. These bounds are combined in the following theorem, as they are 
connected by the duality. 

\begin{theorem}[Rao Bound \cite{rao47} and Hamming (sphere packing) Bound 
\cite{ham50}]
	\index{bound!sphere packing}
	\index{bound!Rao}
	\label{r-h-bound}
	For any code $\mathcal{C} \subseteq F_q^n$ with minimum distance $d$ and 
	dual distance $d^\prime$
	\begin{equation}
		\label{rao-hamming}
		H^{n,q}(d^\prime) \leq |\mathcal{C}| \leq \frac{q^n}{H^{n,q}(d)}.
	\end{equation}
\end{theorem}

\begin{definition}
	\index{code!perfect}
	\index{code!$\tau$-design!tight}
	{\rm Codes attaining the upper bound in \eqref{rao-hamming} are called {\bf 
	perfect}. Designs attaining the lower bound in \eqref{rao-hamming} are 
	called {\bf tight}.}
\end{definition}

Recall the definitions of $s$, $s^\prime$, $\delta$, and $\delta^\prime$ in 
Definition \ref{ddist}.

\begin{theorem}
	\begin{itemize}
		\item[a)] A code $\mathcal{C} \subseteq F_q^n$ ($|\mathcal{C}|>1)$ is a 
		tight design if and only if $d^\prime=2s-\delta+1$. 
		\item[b)] A code $\mathcal{C} \subseteq F_q^n$ ($|\mathcal{C}|>1)$ is a 
		perfect code if and only if $d=2s^\prime-\delta^\prime+1$.
	\end{itemize}
\end{theorem}

See also \cite{banbantanzhu17,dun79,lev99,rao47}.

\begin{definition} 
	{\rm For fixed $n$, $q$, and $i$ denote by $\xi_i^{n,q}$ the smallest root 
	of the Krawtchouk polynomial $K_i^{(n,q)}(z)$. In the case of $i=0$ set 
	$\xi_0^{n,q}=n+1$.}
\end{definition}

\begin{lemma}[\cite{lev95}] 
	\label{k-eps-def}
	The following are valid:
	\begin{itemize}
		\item[a)] For $i=1,2,\ldots,n$, $\xi_i^{n-2,q} < \xi_i^{n-1,q} < 
		\xi_{i-1}^{n-2,q}$. 	 
		\item[b)] For $z \in [1,n]$, there exists a unique $k=k(z) \in 
		\{1,2,\ldots,n\}$ and a unique $\varepsilon=\varepsilon(z) \in \{0,1\}$ 
		such that
		\[ 
			\xi_k^{n-1-\varepsilon,q}+1 < z \leq 
			\xi_{k-1+\varepsilon}^{n-2+\varepsilon,q}+1. 
		\]
	\end{itemize}
\end{lemma}

\begin{example} 
	{\rm One has $\xi_0^{n,q}=n+1$, $\xi_1^{n,q}=\frac{(q-1)n}{q}$ and 
	$\xi_2^{n,q}=\frac{2(q-1)n-q+2-\sqrt{4(q-1)n+(q-2)^2}}{2q}$.}
\end{example}

\begin{definition} 
	{\rm For $z \in [1,n]$, let
	\[ 
		L_k^{n,q}(z) := V_{k-1}(n,q)-(q-1)^k \binom{n}{q} 
		\frac{K_{k-1}^{(n-1,q)}(z-1)}{K_k^{(n,q)}(z)}, \mbox{ and}
	\]
	\[ 
		L^{n,q}(z) := q^{\varepsilon} L_k^{n-\varepsilon,q}(z), \ \mbox{ if } \ 
		\xi_k^{n-1-\varepsilon,q}+1 < z \leq 
		\xi_{k-1+\varepsilon}^{n-2+\varepsilon,q} + 1,
	\]   
	where $k=k(z)$ and $\varepsilon=\varepsilon(z) \in \{0,1\}$ are as in Lemma 
	\ref{k-eps-def}b).   }
\end{definition}

\begin{theorem}[Levenshtein Bound \cite{lev83,lev95,lev98}]
	\label{l-q-bound}
	\index{bound!Levenshtein}
	For any code $\mathcal{C} \subseteq F_q^n$ with minimum distance $d$ and 
	dual distance $d^\prime$
	\begin{equation}
		\label{L-bound}
		\frac{q^n}{L^{n,q}(d^\prime)} \leq |\mathcal{C}| \leq L^{n,q}(d). 
	\end{equation}
	The lower (upper) bound is attained if and only if $d \geq 
	\max\{2s^\prime-\delta^\prime,2\}$ (respectively, $d^\prime \geq 
	\max\{2s-\delta,2\}$).
\end{theorem}

\begin{example}
	\index{bound!Plotkin}
	{\rm In the first three relevant intervals, the Levenshtein (upper) Bound 
	is given by
	\[  
		A_q(n,d) \leq \frac{qd}{qd-(q-1)n}=L_1^{n,q}(d) 
	\]
	(which is the Plotkin Bound; discussed in Section 1.9.3) if $n-\frac{n-1}{q} = \xi_1^{n-1,q} + 1 \leq 
	d \leq \xi_0^{n-2,q}+1=n$,
	\[  
		A_q(n,d) \leq \frac{q^2d}{qd-(q-1)(n-1)} = L_2^{n,q}(d)  
	\]
	if $\xi_1^{n-2,q}+1 \leq d \leq \xi_1^{n-1,q}+1$, and
	\[  
		A_q(n,d) \leq 
		\frac{qd(n(q-1)+1)(n(q-1)-qd+2-q)}{qd(2n(q-1)-q+2-qd)-(n-1)(q-1)^2} = 
		L_3^{n,q}(d)  
	\]
	if $\xi_2^{n-1,q}+1 \leq d \leq \xi_1^{n-2,q}+1$.}
\end{example}

\begin{remark}
	{\rm It is also worth noting two important values of the Levenshtein Bound:
	\[ 			
		L^{n,q}(\xi_{k-1+\varepsilon}^{n-2+\varepsilon,q}+1) = 
		H^{n,q}(2k-1+\varepsilon),  \ \varepsilon \in \{0,1\}, 
	\] 
	i.e. the Levenshtein Bound at the ends of the intervals 
	$\left[\xi_k^{n-1-\varepsilon,q}+1, \xi_{k-1+\varepsilon}^{n-1,q}+1\right]$ 
	coincides with corresponding Rao Bound. }
\end{remark}

Recall the kernels $T_i^{n,q}$ from Definition \ref{Tk-kernels} and the 
parameters $k=k(d)$ and $\varepsilon=\varepsilon(d)$ from Lemma \ref{k-eps-def} 
b). The next theorem gives a Gauss-Jacobi quadrature formula 
\eqref{quad-lev-1}, introduced by Levenshtein in \cite{lev92,lev95}, which is 
instrumental in proofs of Theorem \ref{lev-opt} and Theorem \ref{bdhss-opt} a). 
Theorem \ref{quad-lev} also introduces parameters needed for the universal 
bound \eqref{bdhss-bound} below. 

\begin{theorem}[\cite{lev95}]
	\index{formula!Gauss-Jacobi quadrature}
	\label{quad-lev}
	For any $d \in \{1,2,\ldots,n\}$ the polynomial
	\[ 
		g_d(z) = z(d-z)(n-z)T_{k-1}^{n-1-\varepsilon,q}(z-1,d-1), 
	\]
	where $k=k(d)$ and $\varepsilon=\varepsilon(d) \in \{0,1\}$, has 
	$k+1+\varepsilon$ simple zeros
	\[ 
	\alpha_0 = 0< \alpha_1 = d < \cdots < \alpha_{k+\varepsilon} \leq n 
	\]
	with $\alpha_{k+\varepsilon}=n$ if and only if $\varepsilon=1$ or 
	$\varepsilon=0$ and $d=\xi_{k-1}^{n-2,r}+1$. 

	Moreover, for any real polynomial $f(z)$ of degree at most 
	$2k-1+\varepsilon$ the following equality holds:
	\begin{equation}
		\label{quad-lev-1}
		f_0 = \frac{f(0)}{L_{2k-1+\varepsilon}^{n,q}(d)} + 
		\sum_{i=1}^{k+\varepsilon} \rho_i^{(d)} f(\alpha_i), 
	\end{equation}
	where all coefficients (weights) $\rho_i^{(d)}$, $i=1,2,\ldots,k$, are 
	positive, and, in the case $\varepsilon=1$, $\rho_{k+1}^{(d)} \geq 0$ with 
	equality if and only if $d=\xi_k^{n-1,q}+1$. We have
	\[ 
		\rho_i^{(d)} = \frac{q^{-1-\varepsilon} 
		(q-1)n(n-1)^\varepsilon}{\alpha_i (n-\alpha_i) 
		T_{k-1}^{n-1-\varepsilon,q}(\alpha_i-1, \alpha_i-1)}, 
	\] 
	$i=1,2,\ldots,k$, and 
	\[ 
		\rho_{k+1}^{(d)} = \frac{q^{-\varepsilon} K_k^{(n-\varepsilon,q)}(d)} 
		{K_k^{(n-\varepsilon,q)}(d)K_{k-1}^{(n-1-\varepsilon,q)}(-1) - 
		K_{k-1}^{(n-1-\varepsilon,q)}(d-1)K_k^{(n-\varepsilon,q)}(0)}. 
	\]
\end{theorem}

\begin{remark} 
	{\rm   Levenshtein used the polynomial
	\begin{equation}
		\label{L-poly-hamming}
		f(t)=(t-\alpha_{k+\varepsilon}) (t-\alpha_1)^\varepsilon 
		\prod_{i=1+\varepsilon}^{k-1+\varepsilon} \ (t-\alpha_i)^2 
	\end{equation}
	to obtain the bounds \eqref{L-bound}. It was shown in \cite{boydansto18} 
	that its zeros $\alpha_1,\ldots,\alpha_{k+\varepsilon}$ strongly suggest the
	optimal choice of nodes for the simplex method of Theorem \ref{lp-codes} 
	(equivalently, Theorem 1.9.23 a) from Chapter 1). Computational experiments show that simple replacement of 
	any double zero $\alpha_i \in (j,j+1)$ of Levenshtein's polynomial 
	\eqref{L-poly-hamming} by two simple zeros $j$ and $j+1$ gives in most 
	cases (conjecture: for every sufficiently large rate $d/n$) the best result 
	that can ever be obtained from Theorem \ref{lp-codes}.}
\end{remark}

\begin{remark}
	{\rm  The assertions of Theorem \ref{quad-lev} remain true when $d \in 
	[1,n]$ is a continuous variable. In particular, the quadrature formula 
	\eqref{quad-lev-1} holds true for any real polynomial $f(z)$ of degree at 
	most $2k-1+\varepsilon$ with well defined roots of $g_d(z)$ and 
	corresponding positive $\rho_i^{(d)}$, $i=1,2,\ldots,k+1$.}
\end{remark}

\begin{theorem}[\cite{sid80,lev92}]
	\label{lev-opt}
	The upper bound in (\ref{L-bound}) cannot be improved by a real polynomial 
	$f$ of degree at most $2k-1+\varepsilon$ satisfying (A1) and the condition 
	$f(z) \leq 0$ for $z \in [d,n]$.
\end{theorem}

\begin{definition}[Test functions \cite{boydan98}]
	\index{test functions}
	{\rm  For $j \in \{1,2,\ldots,n\}$ and $d \in (0,n]$ denote by
	\[ 
		P_j^{n,q}(d) := \frac{K_j^{(n,q)}(0)}{L_{2k-1+\varepsilon}^{n,q}(d)} + 
		\sum_{i=1}^{k+\varepsilon} \rho_i^{(d)} K_j^{(n,q)}(\alpha_i), 
	\]
	where $k=k(d)$, $\varepsilon=\varepsilon(d)$, and $\rho_i^{(d)}$ are as in 
	Theorem \ref{quad-lev}. Note that $P_j^{n,q}=0$ for $j \leq 2k-1 + 
	\varepsilon$. Note also that $K_j^{(n,q)}(0)=(q-1)^j \binom{n}{j}$.}
\end{definition}

\begin{theorem} [\cite{boydan98,lev98}]
	The upper bound in \eqref{L-bound} can be improved by a real polynomial $f$ 
	of degree at least $2k+\varepsilon$ satisfying (A1) and the condition $f(z) 
	\leq 0$ for $z \in [d,n]$  if and only if $P_j^{n,q}(d) < 0$ for some $j 
	\geq 2k+\varepsilon$. 
\end{theorem}

\begin{definition}
	\index{function!completely monotone}
	{\rm A function $h:(0,n] \to (0,+\infty)$ is called {\bf (strictly) 
	completely monotone} if $(-1)^i h^{(i)}(z) \geq 0$ ($>0$) for every 
	nonnegative integer $i$ and every $z \in (0,n]$. The derivatives 
	$h^{(i)}(z)$ can be discrete (then $z \in \{1,2,\ldots,n\}$) or continuous.}
\end{definition}

\begin{theorem}[Universal lower bound on energy \cite{boydraharsafsto17}]
	\index{bound!on code energy}
	Let $n$, $q$ and $M \in \{2,3,\ldots,q^n\}$ be fixed and $h:(0,n] \to 
	(0,+\infty)$ be a completely monotone function. If $M \in 
	\left(H^{n,q}(2k-1+\varepsilon), H^{n,q}(2k+\varepsilon)\right]$ and 
	$d=d(M) \in [1,n]$ is the smallest real root of the equation 
	$M=L_{2k-1+\varepsilon}^{n,q}(d)$, then
	\begin{equation}
	\label{bdhss-bound}
		E_h(n,M;q) \geq M^2 \sum_{i=1}^{k+\varepsilon} \rho_i^{(d)} 
		h(\alpha_i), 
	\end{equation}
	where the parameters $\alpha_i$ and $\rho_i^{(d)}$ are determined as in 
	Theorem \ref{quad-lev}. Equality is attained if and only if there exists a 
	code with $M$ codewords which attains the upper bound in \eqref{L-bound}. 
\end{theorem}

\begin{theorem}[\cite{boydraharsafsto17}]
	\label{bdhss-opt}
	Let $h$ be a strictly absolutely monotone function. The bound 
	\eqref{bdhss-bound}:
	\begin{itemize}
		\item[a)] cannot be improved by a real polynomial $f$ of degree at most 
		$2k-1+\varepsilon$ satisfying (A1) and $f(z) \leq h(z)$ for $z \in 
		[d,n]$;
		\item[b)] can be improved by a real polynomial $f$ of degree at least 
		$2k+\varepsilon$ satisfying (A1) and the condition  $f(z) \leq h(z)$ 
		for $z \in [d,n]$  if and only if $P_j^{n,r}(d) < 0$ for some $j \geq 
		2k + \varepsilon$.
	\end{itemize}
\end{theorem}

\begin{definition}
	\index{code!optimal!universally}
	{\rm A code $\mathcal{C} \subseteq F_q^n$ is called {\bf universally 
	optimal} if it (weakly) minimizes potential energy among all configurations 
	of $|\mathcal{C}|$ codewords in $F_q^n$ for each completely monotone 
	function $h$.}
\end{definition}

The conditions for attaining the bounds \eqref{L-bound} and \eqref{bdhss-bound} 
coincide. Thus, any code which attains the upper bound in \eqref{L-bound} for 
its cardinality (and therefore the lower bound for \eqref{bdhss-bound} for its 
energy) is universally optimal. We summarize this in the next theorem. 

\begin{theorem}[\cite{cohzha14,boydraharsafsto17}]
	All codes which attain the upper bound in \eqref{L-bound} are universally 
	optimal. 
\end{theorem}

\begin{theorem}[\cite{cohzha14}]
	Let $\mathcal{C} \subseteq F_q^n$ be a code and $h:(0,1,\ldots,n] \to 
	\mathbb{R}$ be any function such that the bound in Theorem 
	\ref{LP-hamming-energy} is attained. Let $c \in \mathcal{C}$. Then 
	\[ 
		E_h(\mathcal{C} \setminus \{c\}) = E_h(n,|\mathcal{C}|-1;q). 
	\]
	In particular, if $\mathcal{C}$ is proved to be universally optimal by 
	Theorem  \ref{LP-hamming-energy}, then $\mathcal{C} \setminus \{c\}$ is 
	universally optimal for all $c \in \mathcal{C}$ as well. 
\end{theorem}

\begin{table}
	\index{code!optimal!table}
	\begin{center} 	
	   \begin{tabular}{|c|c|c|c|c|c|p{90pt}|}
		\hline
		$n$ & $q$ & $s'(\mathcal{C})$ & $\!d'(\mathcal{C})\!$ & Distances & 
		$|\mathcal{C}|$ & Comment \\ 
		\hline
		$n$ & $q$ & $n$ & $2$ & $\{n\}$ & $n$ & 
		repetition code, $s'(\mathcal{C})=\lfloor n/2\rfloor$ for $q=2$\\ 
		\hline
		$n$ & $q$ & $n-1$ & $2$ & $\{d\}$ & $\frac{qd}{qd-n(q-1)}$ & 
		$n>d>\frac{(q-1)n+1}{q}$ coexistence  with 
		resolvable block-designs 
		$2-(|\mathcal{C}|,\frac{|\mathcal{C}|}{q},n-d)$ \cite{semzin68}\\
		\hline 
		$n$ & $q$ & $n-2$ & $3$ & $\{\frac{(q\!-\!1)n\!+\!1}{q}\}$ & $(q-1)n+1$ 
		& 
		coexistence with affine resolvable block-designs 
		$2-(|\mathcal{C}|,\frac{|\mathcal{C}|}{q},\frac{n-1}{q})$ 
		\cite{semzinzai69} \\ 
		\hline 
		$p^lq$ & $\!q\!=\!p^m\!$ & $n-2$ & $3$ & $\{n-p^l,n\}$ & $nq$ & 
		$l,m=1,2,\dots$, \cite{semzinzai69} \\ 
		\hline
		$\!\!qh\!+\!h\!-\!q\!\!$ & $q$ & $n-2$ & $3$ & $\{n-h,n\}$ & $q^3$ & 
		$2\vert q, h\vert q, 2<h<q$, \cite{den69} \\ 
		\hline 
		$q^2+1$ & $q$ & $n-3$ & $4$ & $\{q^2-q,q^2\}$ & $q^4$ & 
		ovoid in PG$(3,q)$, \cite{bos47,qvi52} \\ 
		\hline 
		$56$ & $3$ & $53$ & $4$ & $\{36,45\}$ & $3^6$ & 
		Projective cap, Hill \cite{hil76} \\ 
		\hline
		$78$ & $4$ & $75$ & $4$ & $\{56,64\}$ & $4^6$ & 
		Projective cap, Hill \cite{hil76} \\ 
		\hline
		$4l$ & $2$ & $2l-2$ & $4$ & $\{2l,4l\}$ & $8l$ & 
		Hadamard codes\\ 
		\hline
		$q+2$ & $q$ & $q-1$ & $4$ & $\{q,q+2\}$ & $q^3$ & 
		$2\vert q$, $s'(\mathcal{C})=2$ for $q=4$; hyperoval in PG$(2,q)$, \cite{bos47} \\ 
		\hline
		$11$ & $3$ & $5$ & $5$ & $\{6,9\}$ & $243$ & 
		projection of Golay code \\ 
		\hline
		$12$ & $3$ & $3$ & $6$ & $\{6,9,12\}$ & $729$ & 
		Golay code \cite{gol49} \\ 
		\hline
		$22$ & $2$ & $10$ & $6$ & $\{8,12,16\}$ & $1024$ & 
		projection of Golay code\\ 
		\hline
		$23$ & $2$ & $7$ & $7$ & $\{8,12,16\}$ & $2048$ & 
		projection of Golay code\\ 
		\hline
		$24$ & $2$ & $4$ & $8$ & $\{8,12,16,24\}$ & $4096$ & 
		Golay code \cite{gol49} \\ 
		\hline
		$n$ & $2$ & $1$ & $n$ & all even & $2^{n-1}$ & 
		even weight code\\
		\hline 
	   \end{tabular}
	\end{center}
	\caption{
		Parameters of known codes attaining (simultaneously) the upper bound in 
		\eqref{L-bound} (see also \cite{boydan97}) and the lower bound 
		\eqref{bdhss-bound}. This table appears for \eqref{L-bound} in 
		\cite{lev92}. Here the column for the external distance $s^\prime$ is 
		added. All codes in the table are universally optimal. 
	}
	\label{Table1-lev-rao}
\end{table}

\begin{example} 
	\index{code!Kerdock $\mathcal{K}_m$}
	\label{Ker-Ham} 
	{\rm The Kerdock codes $\mathcal{K}_{\ell} \subset 
	\mathbb{F}_2^{2^{2\ell}}$ \cite{Ker72} are nonlinear codes existing for 
	lengths $n=2^{2\ell}$. Their cardinality is 
	$|\mathcal{K}_\ell|=n^2=2^{4\ell}$ and their distance (weight) distribution 
	is as follows:
	\[ 
		B_{2^{2\ell-1}-2^{\ell-1}} = 2^{2\ell}(2^{2\ell-1}-1), \ 
		B_{2^{2\ell-1}} = 2^{2\ell+1}-2, \ B_{2^{2\ell-1}+2^{\ell-1}} = 
		2^{2\ell}(2^{2\ell-1}-1), 
	\]
	\[ 
		B_0 = B_n =1, \ B_i = 0 \ \mbox{for } i \neq 0, 2^{2\ell-1}-2^{\ell-1}, 
		2^{2\ell-1}, 2^{2\ell-1}+2^{\ell-1}. 
	\]
	It is easy to check that the Kerdock codes are asymptotically optimal with 
	respect to the upper bound in \eqref{L-bound} and the bound 
	\eqref{bdhss-bound} as they are very close to the bounds already for small 
	$\ell$.}
\end{example}

\begin{definition}
	\index{code!covering radius}
	{\rm   For any code $\mathcal{C} \subseteq F_q^n$, $\rho(\mathcal{C}):= 
	\max\{d(x,\mathcal{C}) :  x \in F_q^n \}$ is called the {\bf covering 
	radius} of $\mathcal{C}$.  Here $d(x,\mathcal{C}):=\min\{d(x,c): c \in 
	\mathcal{C}\}$.}
\end{definition}

\begin{theorem}[\cite{del73a}, Delsarte Bound]
	\index{bound!Delsarte}
	For any code $\mathcal{C} \subseteq F_q^n$ with external distance $s^\prime$
	\[ 
		\rho(\mathcal{C}) \leq s^\prime. 
	\]   
\end{theorem}

\begin{theorem} [\cite{tie90,tie91}, Tiet\"av\"ainen Bound]
	\index{bound!Tiet\"av\"ainen}
	For any code $\mathcal{C} \subseteq F_q^n$ with dual distance $d^\prime = 
	2k - 1 + \varepsilon$, where $k$ is positive integer and $\varepsilon \in 
	\{0,1\}$,
	\[ 
		\rho(\mathcal{C}) \leq \xi_k^{n-1+\varepsilon,q}. 
	\]
\end{theorem}

See also \cite{fazlev95,sol90}.

\noindent
Asymptotic versions of some of the bounds in this section can be found in 
Section 1.9.8 of Chapter 1 (see \cite{aal79,aal90,barjaf01,lev98,mcerodrumwel77}).

\section{Linear programming on $\mathbb{S}^{n-1}$}
\label{s4}

\begin{definition}
	\index{inner product!Euclidean}
	\index{distance!Euclidean}
	\label{SphericalDistInnPr}
	{\rm Let $\mathbb{S}^{n-1}=\{{\bf x}=(x_1,x_2,\dots,x_n): 
	x_1^2+x_2^2+\cdots+x_n^2=1\}$ be the unit sphere in $\mathbb{R}^n$. The 
	{\bf Euclidean distance} between ${\bf x}=(x_1,x_2,\dots,x_n)$ and 
	${\bf y}=(y_1,y_2,\dots,y_n)$ is
	\[
		d({\bf x},{\bf y}) := \sqrt{(x_1-y_1)^2+(x_2-y_2)^2+\cdots + 
		(x_n-y_n)^2}.
	\]
	The {\bf inner product}  is defined as
	\[
		({\bf x},{\bf y}) : =x_1y_1+x_2y_2+\dots +x_ny_n.
	\]}
\end{definition}

Note that on $\mathbb{S}^{n-1}$ the distance and the inner product are connected by
\[ ({\bf x},{\bf y})=1-\frac{d^2({\bf x},{\bf y})}{2}. \]

\begin{definition}
	\index{code!spherical}
	\label{SphericalCode}
	{\rm An $(n,M,s)${\bf{}-spherical code} is a non-empty finite set 
	$\mathcal{C} \subset \mathbb{S}^{n-1}$ with cardinality $|\mathcal{C}|=M$ 
	and maximal inner product 
	\[ 
		s=\max\{ ({\bf x},{\bf y}) : {\bf x},{\bf y} \in \mathcal{C}, {\bf x} 
		\neq {\bf y}\}.
	\]}
\end{definition}

The minimum distance $d=d(C):=\min \{d({\bf x},{\bf y}):  {\bf x},{\bf y} \in \mathcal{C}, {\bf x} \neq {\bf y}\}$ and the 
maximal inner product are connected by 
\[ s=1-\frac{d^2}{2}. \]

\begin{definition}
	\index{code!optimal!$A(n,s)$}
	{\rm   For fixed $n$ and  $s$ denote by
	\[ 
		A(n,s) := \max \{ |\mathcal{C}|: \mathcal{C} \mbox{ is an } 
		(n,M,s)\mbox{-spherical code}\}.
	\]}
\end{definition}

\begin{definition}
	\index{polynomial!Jacobi}
	\index{polynomial!Gegenbauer}
	{\rm    For $a,b \in \{0,1\}$ denote by $\{P_i^{a,b}(t)\}_{i=0}^\infty$ the 
	Jacobi polynomials $\{P_i^{\alpha,\beta}(t)\}_{i=0}^\infty$ (\cite[Chapter 
	22]{abrste64}, \cite{sze39}) with
	\[ 
		(\alpha,\beta) = \left( a+\frac{n-3}{2}, b+\frac{n-3}{2}\right) 
	\]
	normalized by  $P_i^{\alpha,\beta}(1)=1$. When $(a,b)=(0,0)$ we get the 
	Gegenbauer polynomials and use the $(n)$ indexing instead of $0,0$. Denote by $t_i^{a,b}$ the 
	greatest zero of the polynomial $P_i^{a,b}(t)$ and define $t_0^{1,1}=-1$. Let 
	\[ r_i:=\frac{2i+n-2}{i+n-2} \binom{i+n-2}{i}. \] }
\end{definition}

\begin{theorem}[\cite{abrste64,sze39}]
	\label{geg-1}
	\index{polynomial!Gegenbauer!three-term reccurence}
	The Gegenbauer polynomials $\{P_i^{(n)}(t)\}_{i=0}^\infty$ satisfy the 
	recurrence relations
	\[ 
		(i+n-2)P_{i+1}^{(n)}(t) = (2i+n-2)tP_i^{(n)}(t) - iP_{i-1}^{(n)}(t),\ \ 
		i=1,2,\ldots ,
	\]
	where  $P_0^{(n)}(t)=1$ and $P_1^{(n)}(t)=t$. 
\end{theorem}

\begin{theorem}[\cite{abrste64,sze39}]
	\index{polynomial!orthogonality}
	\label{geg-2}
	The Gegenbauer polynomials are orthogonal on $[-1,1]$ with respect to the 
	measure 
	\[
		d\mu(t) := \gamma_n (1-t^2)^{\frac{n-3}{2}}\, dt, \quad t\in [-1,1], \]
	where $\gamma_n := \Gamma(\frac{n}{2})/\sqrt{\pi}\Gamma(\frac{n-1}{2})$ is a normalizing constant.  
\end{theorem}

\begin{theorem}[Linear Programming Bound for spherical codes, 	    
\cite{delgoesei77,kablev78}] 
	\index{bound!linear programming!spherical codes}
	\label{lp-max-sc}
	Let $n \geq 2$ and $f(t)$ be a real polynomial such that
	\begin{itemize}
		\item[(A1)] $f(t) \leq 0 $ for $-1 \leq t \leq s$;
		\item[(A2)] The coefficients in the Gegenbauer expansion $f(t)= 
		\sum_{i=0}^{\deg(f)} f_i P_i^{(n)}(t)$ satisfy $f_0>0$, $f_i \geq 0$ 
		for $i=1,\ldots,\deg(f)$.
	\end{itemize}
	Then  $A(n,s) \leq f(1)/f_0$.
\end{theorem}

\begin{theorem}[Levenstein Bound \cite{lev79,lev92}]
	\index{bound!Levenshtein}
	For the quantity $A(n,s)$ we have
	\begin{equation}
		\label{L-bound-SC}
		A(n,s) \leq L_{\tau}(n,s):= \left(1-\frac{P_{k-1+\varepsilon}^{1,0}(s)} 
		{P_k^{0,\varepsilon}(s)}\right) \sum_{i=0}^{k-1+\varepsilon} r_i,  \ 
		\forall s \in I_{\tau},     
	\end{equation}
	where $I_\tau$ is the interval 
	\[ 
		I_\tau : =\left[t_{k-1+\varepsilon}^{1,1-\varepsilon}, 
		t_k^{1,\varepsilon}\right], \ \ \tau=2k-1+\varepsilon, \ \varepsilon 
		\in \{0,1\}.
	\]
\end{theorem}

\begin{example} 
	{\rm The first three bounds in \eqref{L-bound-SC} are $A(n,s) \leq (s-1)/s$ 
	for $s \in [-1,-1/n]$,
	\[ 
		A(n,s) \leq \frac{2n(1-s)}{1-ns} 
	\]
	for $s \in [-1/n,0]$ and
	\[ 
		A(n,s) \leq \frac{n(1-s)(2+(n+1)s)}{1-ns^2} 
	\]
	for $s \in \left[0,\frac{\sqrt{n+3}-1}{n+2}\right]$.}
\end{example}

See also \cite{erizin01,lev83,lev98}.

\begin{definition}
	\index{code!$\tau$-design!spherical}
	{\rm A spherical code $\mathcal{C} \subset \mathbb{S}^{n-1}$ is a {\bf 
	spherical $\tau$-design} if and only if 
	\[ 
		\int_{\mathbb{S}^{n-1}} p({\bf x}) d\sigma_n({\bf x}) = 
		\frac{1}{|\mathcal{C} |} \sum_{{\bf x} \in \mathcal{C} } p({\bf x)} 
	\]
	($\sigma_n $ is the normalized $(n-1)$-dimensional Hausdorff measure) holds 
	for all polynomials $p({\bf x}) = p(x_1,x_2,\ldots,x_n)$ of degree at most 
	$\tau$.}
\end{definition}

\begin{theorem}[Linear Programming Bound for spherical designs, 	
\cite{delgoesei77}] 
	\index{bound!linear programming!spherical designs}
	Let $n \geq 2$, $\tau \geq 1$ and $f(t)$ be a real polynomial such that
	\begin{itemize}
		\item[(B1)] $f(t) \geq 0 $ for $-1 \leq t \leq 1$;
		\item[(B2)] the coefficients in the Gegenbauer expansion $f(t)= 
		\sum_{i=0}^{\deg(f)} f_i P_i^{(n)}(t)$ satisfy $f_0>0$, $f_i \leq 0$ 
		for $i=\tau+1,\ldots,\deg(f)$.
	\end{itemize}
	Then any spherical $\tau$-design $\mathcal{C} \subset \mathbb{S}^{n-1}$ has 
	cardinality $|\mathcal{C}| \geq f(1)/f_0$.
\end{theorem}

\begin{theorem}[Delsarte-Goethals-Seidel Bound \cite{delgoesei77}]
	\index{bound!Delsarte-Goethals-Seidel}
	Any $\tau$-design $\mathcal{C} \subseteq \mathbb{S}^{n-1}$ has cardinality
	\begin{equation}
		\label{DGS-bound}
		|\mathcal{C}| \geq D(n,\tau) := \binom{n+k-2+\varepsilon}{n-1} + 
		\binom{n+k-2}{n-1},
	\end{equation}
	where $\tau=2k-1+\varepsilon$, $\varepsilon \in \{0,1\}$. 
\end{theorem}

\begin{definition} 
	\index{code!$\tau$-design!tight}
	{\rm  A spherical $\tau$-design on $\mathbb{S}^{n-1}$ is called {\bf tight} 
	if it attains the bound \eqref{DGS-bound}.}
\end{definition}

\begin{theorem}[\cite{BanDam79,BanDam80}] 
	Let $n \geq 3$. Tight spherical $\tau$-designs on $\mathbb{S}^{n-1}$ exist 
	for $\tau=1$, $2$ and $3$ for every $n \geq 2$, and possibly for $\tau=4$, 
	$5$, $7$, and $11$. Tight spherical $4$-designs on $\mathbb{S}^{n-1}$ exist for $n=6$, $22$, and 
possibly for $m^2-3$, where $m \geq 7$ is an odd integer. Tight 
	spherical $5$-designs on $\mathbb{S}^{n-1}$ exist for $n=3$, $7$, $23$, and possibly 
for $n=m^2-2$, where $m \geq 7$ is an odd integer. Tight 
	spherical $7$-designs on $\mathbb{S}^{n-1}$ exist for $n=8$, $23$, and possibly for 
	$n=3m^2-4$, where $m \geq 4$ is an integer. Tight spherical $11$-designs on 
	$\mathbb{S}^{n-1}$ exist only for $n=24$.  
\end{theorem}

\begin{remark}[\cite{delgoesei77}]
	{\rm  Tight spherical 4- and 5-designs coexist and are known for $m=3$ and 
	$5$ only. Tight spherical 7-designs are known for $m=2$ and $3$ only. }
\end{remark}

See also \cite{banban09,banbantanzhu17,bonradvia13} for general theory and 
\cite{bouboydan99,fazlev95,sol91} for other bounds for designs.

\begin{theorem}[\cite{lev79,lev83,lev92,lev98}]
	The bounds and \eqref{L-bound-SC} and \eqref{DGS-bound} are related by the 
	equalities
	\begin{equation}
		\label{L-DGS1}
		L_{\tau-1-\varepsilon}(n,t_{k-1-\varepsilon}^{1,1-\varepsilon}) = 
		L_{\tau-\varepsilon}(n,t_{k-1-\varepsilon}^{1,1-\varepsilon}) = 
		D(n,\tau-\varepsilon), \ \varepsilon \in \{0,1\}
	\end{equation}
	at the ends of the intervals $I_\tau$ ($\tau=2k-1+\varepsilon$, 
	$\varepsilon \in \{0,1\}$). In particular, if $\mathcal{C} \subseteq 
	\mathbb{S}^{n-1}$ is a tight spherical $\tau$-design, then it attains 
	(\ref{L-bound-SC}) in the left end of the interval $I_\tau$. 
\end{theorem}

\begin{definition}
	\index{code!potential energy}
	{\rm   Given an (extended real-valued) function $h(t):[-1,1] \to 
	[0,+\infty]$, the $h${\bf -energy} (or {\bf potential energy}) of 
	$\mathcal{C}$ is given by
	\[ 
		E_{h}(\mathcal{C}) := \sum_{{\bf x}, {\bf y} \in \mathcal{C}, {\bf x} 
		\neq {\bf y}} h(({\bf x}, {\bf y})). 
	\]}
\end{definition}

\begin{definition} 
	\index{code!optimal!$E_h(n,M)$}
	{\rm  For fixed $n$, $h$, and $M \geq 2$ denote by
	\[  
		\mathcal{E}_h(n,M) := \inf \{E_h(\mathcal{C}): \mathcal{C} \subset 
		\mathbb{S}^{n-1}, \ |\mathcal{C}|=M\}. 
	\]}
\end{definition}

\begin{theorem}[Linear Programming Bound for energy \cite{yud92}]
	\index{bound!on code energy}
	\label{pl-energy-sc}
	Let $n \geq 2$, $M \geq 2$,  $h:[-1,1]\to[0,+\infty]$, and $f(t)$ be a real 
	polynomial such that
	\begin{itemize}
		\item[(C1)] $f(t) \leq h(t)$ for $t\in [-1,1]$, and
		\item[(C2)] the coefficients in the Gegenbauer expansion $f(t)= 
		\sum_{i=0}^{\deg(f)} f_i P_i^{(n)}(t)$ satisfy $f_0>0$, $f_i \geq 0$ 
		for $i=1,\ldots,\deg(f)$.
	\end{itemize}
	Then 
	\[ 
		\mathcal{E}_h(n,M) \geq M(f_0M-f(1)). 
	\]
\end{theorem}

\begin{definition}
	\index{function!absolutely monotone}
	{\rm    A real valued extended function $h(t):[-1,1] \to (0,+\infty]$ is 
	called {\bf absolutely monotone} if $h^{(k)}(t)\geq 0$, for every 
	$t\in[-1,1)$ and every integer $k \geq 0$, and $h(1)=\lim_{t \to 1^{-}} 
	h(t)$.}
\end{definition}

The following absolutely monotone potential functions are commonly used: 
Newton potential 
\[ h(t)=(2-2t)^{-(n-2)/2}=d({\bf x},{\bf y})^{-(n-2)}, \]
Riesz $s$-potential 
\[ h(t)=(2-2t)^{-s/2}=d({\bf x},{\bf y})^{-s}, \]
and Gaussian potential 
\[ h(t)=\exp(2t-2)=\exp(-d({\bf x},{\bf y})^2). \]

\begin{definition} 
	\label{defn-alfa-ro} 
	{\rm For $(a,b)=(0,0), (1,0),$ and $(1,1)$ set
	\[ 
		T_j^{a,b}(u,v) := \sum_{i=0}^j r_i^{a,b} P_i^{a,b}(u)P_i^{a,b}(v), 
	\]
	where $r_i^{a,b}=1/c^{a,b}\int_{-1}^1 
	\left(P_i^{a,b}(t)\right)^2(1-t)^a(1+t)^b d \mu(t)$, 
	$c^{1,0}=c^{0,0}=\gamma_n$, and $c^{1,1}=\gamma_{n+2}$ (see Theorem 
	\ref{geg-2} for relevant notation). Note that $r_i^{0,0}=r_i$. 

Let $\alpha_1 < \alpha_2 < \cdots < 
	\alpha_{k+\varepsilon}$ be the roots of the polynomial used for obtaining 
	the Levenshtein Bound $L_\tau(n,s)$ with $s=\alpha_{k+\varepsilon}$, 
	$\tau=2k-1+\varepsilon$, $\varepsilon \in \{0,1\}$, and let
	\[ 
		\rho_1 := \frac{T_k^{0,0}(s,1)} 
		{T_k^{0,0}(-1,-1)T_k^{0,0}(s,1)-T_k^{0,0}(-1,1)T_k^{0,0}(s,-1)} \ \ for 
		\ \ \varepsilon=1, 
	\]
	\[ 
		\rho_{i+\varepsilon} := \frac{1}{c^{1,\varepsilon}(1 + 
		\alpha_{i+\varepsilon})^\varepsilon(1-\alpha_{i+\varepsilon}) 
		T_{k-1}^{1,\varepsilon}(\alpha_{i+\varepsilon},\alpha_{i+\varepsilon})},
		 \ \ i=1,2,\ldots,k.
	\]}
\end{definition} 

\begin{theorem}[\cite{boydraharsafsto16}]
	Let $n \geq 2$, $\tau =2k-1+\varepsilon \geq 1$, $\varepsilon \in \{0,1\}$, 
	and $h$ be absolutely monotone in $[-1,1]$. For $M \in 
	\left(D(n,\tau),D(n,\tau+1)\right]$ let $s$ be the largest root of the 
	equation $M=L_\tau(n,s)$ and $\alpha_1 < \alpha_2 < \cdots < 
	\alpha_{k+\varepsilon} = s$, $\rho_1,\rho_2,\ldots,\rho_{k+\varepsilon}$ be 
	as in Definition \ref{defn-alfa-ro}. Then 
	\begin{equation}
		\label{ULB1}
		\mathcal{E}_h(n,M) \ge M^2\sum_{i=1}^{k+\varepsilon} \rho_i h(\alpha_i).
	\end{equation}
	If an $(n,M,s)$ code attains \eqref{ULB1}, then it is a spherical 
	$\tau$-design and its inner products form the set 
	$\{\alpha_1,\alpha_2,\ldots,\alpha_{k+\varepsilon}\}$.   
\end{theorem}

See also \cite[Chapter 5]{borharsaf18}. Note that $\alpha_1, \alpha_2, \cdots, 
\alpha_{k+\varepsilon} = s$ are in fact the roots of the equation 
$M=L_\tau(n,s)$.

\begin{remark} 
	{\rm The conditions for attaining the bounds  \eqref{L-bound-SC} and 
	\eqref{ULB1} coincide. Thus a spherical $(n,M,s)$ code attains 
	\eqref{L-bound-SC} if and only it attains \eqref{ULB1}. In particular, 
	every tight spherical design attains \eqref{ULB1}.   }
\end{remark}

\begin{theorem}[\cite{sid80} for \eqref{L-bound-SC}, \cite{boydraharsafsto16} 
for \eqref{ULB1}] 
	The bounds \eqref{L-bound-SC} and \eqref{ULB1} cannot be improved by using 
	in Theorem \ref{lp-max-sc} and \ref{pl-energy-sc}, respectively, 
	polynomials of the same or lower degree.
\end{theorem}

\begin{theorem}[\cite{boydanbum96} for  \eqref{L-bound-SC}, 
\cite{boydraharsafsto16} for \eqref{ULB1}]
	Let 
	\[ 
		Q_j^{(n)}:= \frac{1}{L_\tau(n,\alpha_{k+\varepsilon})} + 
		\sum_{i=1}^{k+\varepsilon}\rho_i P_j^{(n)}(\alpha_i), \ \ 
		j>\tau=2k-1+\varepsilon, \ \varepsilon \in \{0,1\}. 
	\]
	The bounds  \eqref{L-bound-SC} and \eqref{ULB1} (where $M = 
	L_\tau(n,\alpha_{k+\varepsilon})$ for  \eqref{ULB1}) can be 
	(simultaneously) improved if and only if $Q_j^{(n)}<0 $ for some $j>\tau$. 
\end{theorem}

\begin{definition}
	\index{code!optimal!universally}
	{\rm    A spherical code $\mathcal{C} \subset \mathbb{S}^{n-1}$ is {\bf 
	universally optimal} if it (weakly) minimizes $h$-energy among all 
	configurations of $|\mathcal{C}|$ points on $S^{n-1}$ for each absolutely 
	monotone function $h$.}
\end{definition}

\begin{theorem}[\cite{cohkum07}]
	Every spherical code which is a spherical $(2k-1)$-design and which admits 
	exactly $k$ inner products between distinct points is universally optimal. 
	The 600-cell (the unique $(4,120,(1+\sqrt{5})/4)$-spherical code) is 
	universally optimal.
\end{theorem}

See also \cite{balblecohgiakelshu09,borharsaf18,cohwoo12}. 

\begin{remark}
	{\rm Any code which attains \eqref{L-bound-SC} (and \eqref{ULB1}) is 
	universally optimal.  Is it unknown if there exists a spherical code, apart 
	from the 600-cell, which is universally optimal but does not attain 
	\eqref{L-bound-SC} (and \eqref{ULB1}). Tables with all known codes which 
	attain  \eqref{L-bound-SC} (and \eqref{ULB1}) can be found in 
	\cite{lev92,lev98,cohkum07}.}
\end{remark}

\begin{example}
	{\rm    Consider the case $(n,M)=(4,24)$. The well known code $D_4$ ($D_4$ 
	root system; equivalently, the set of vertices of the regular $24$-cell) is 
	optimal in the sense that it realizes the fourth kissing number 
	\cite{mus08}. However, this code is not universally optimal 
	\cite{cohconelkkum07}, despite having energy which is very close to the 
	bound \eqref{ULB1}. For example, with the Newtonian $h(t)=\frac{1}{2(1-t)}$ it has energy 
	$334$, while \eqref{ULB1} gives $333$ (which can be improved to $\approx 
	333.157$).}
\end{example}

\begin{conjecture}
	{\rm Every universally optimal spherical code attains the Linear 
	Programming Bound of Theorem \ref{pl-energy-sc}.  }
\end{conjecture}

\begin{example} 
	\index{code!Kerdock $\mathcal{K}_m$}
	\label{Ker-Sph} 
	{\rm (Example \ref{Ker-Ham} continued.) A standard construction (see 
	\cite[Chapter 5]{ConSlo88}) maps binary codes from $\mathbb{F}_2^n$ to the 
	sphere $\mathbb{S}^{n-1}$ -- the coordinates $0$ and $1$ are replaced by 
	$\pm 1/\sqrt{n}$, respectively. Denote this map by ${\bf x} \to 
	\overline{{\bf x}}$. The inner product $( \overline{{\bf x}},\overline{{\bf 
	y}})$ on $\mathbb{S}^{n-1}$ and the Hamming distance ${\rm d_H}({\bf 
	x},{\bf y})$ in $\mathbb{F}_2^n$ are connected by $(\overline{{\bf 
	x}},\overline{{\bf y}}) = 1-\frac{2{\rm d_H}({\bf x},{\bf y})}{n}$. Thus 
	the weights of the Kerdock code $\mathcal{K}_\ell$ correspond to the inner 
	products $1,\frac{1}{\sqrt{n}},0,-\frac{1}{\sqrt{n}},-1$, respectively.

	The image $\overline{\mathcal{K}}_{\ell} \subset \mathbb{S}^{2^{2\ell}-1}$ 
	of $\mathcal{K}_\ell$ is asymptotically optimal with respect to both bounds 
	\eqref{L-bound-SC} and \eqref{ULB1}. For example, it has energy
	\[ 
		E_h(\overline{\mathcal{K}}_{\ell}) = n^2 
		\left(\left(2^{2\ell+1}-2\right)h(0) + 
		2^{2\ell}(2^{2\ell-1}-1)\left(h\left(\frac{1}{2^\ell}\right) + 
		h\left(-\frac{1}{2^\ell}\right)\right)+h(-1)\right). 
	\]
	When $n$ tends to infinity, we obtain
	\[  
		E_h(\overline{\mathcal{K}}_{\ell}) \sim n^2\left( (2^{4\ell}-2) 
		h(0)+h(-1)\right) \sim h(0)n^4, 
	\]
	which coincides with the asymptotic of \eqref{ULB1}  (obtained by a 
	polynomial of degree 5).}
\end{example}

\section{Linear programming in other coding theory problems}
\label{LPB:s5}

\begin{remark}[Linear programming in Johnson spaces]
	\index{bound!linear programming!Johnson spaces}
	{\rm   All concepts and results from Sections \ref{LPB:s1}-\ref{LPB:s3} 
	hold true for the Johnson spaces with changes corresponding to the role of 
	the Hahn polynomials instead of the Krawtchouk polynomials. See 
	\cite{del73b,dellev98,lev92,lev98,lev99,mcerodrumwel77}.}
\end{remark}

\begin{theorem}[Linear Programming Bound for binomial moments \cite{ashbar99}]
	\index{bound!linear programming!binomial moments}
	Let $\mathcal{C} \subset F_2^n$ be a code with distance distribution 
	$(B_0,B_1,\ldots,B_n)$, $w \in \{1,2,\ldots,n\}$, and
	\[ 
		\mathcal{B}_w := \sum_{i=1}^w \binom{n-i}{n-w} B_i .
	\]
	Let the real polynomial $f(z)=\sum_{i=0}^n f_i K_i^{(n,2)}(z)$ satisfy the 
	conditions
	\begin{itemize}
		\item[(i)] $f_i \geq 0$ for $i=1,2,\ldots,n$;
		\item[(ii)] $f(j) \leq \binom{n-j}{n-w}$ for $j=1,2,\ldots,n$.
	\end{itemize}
	Then $\mathcal{B}_w \geq f_0|\mathcal{C}|-f(0)$. 
\end{theorem}

The final bound in this chapter is an upper bound on the size of a quantum 
code.  

\begin{theorem}[Linear Programming Bound for quantum codes \cite{ashlit99}] 
(see also \cite{laiash18})
	\index{bound!linear programming!quantum codes}
	Let $f(z)=\sum_{i=0}^n f_i K_i^{(n,4)}(z)$ be a polynomial satisfying the 
	conditions
	\begin{itemize}
		\item[(i)] $f_i \geq 0$ for every $i=0,1,\ldots,n$;
		\item[(ii)] $f(z) > 0$ for every $z=0,1,\ldots,d-1$;
		\item[(iii)] $f(z) \leq 0$ for every $z=d,d+1,\ldots,n$.
	\end{itemize}
	Then every $((n,K))$ quantum code of minimum distance $d$ satisfies
	\[ 
		K \leq \frac{1}{2^n} \max_{j \in \{0,1,\ldots,d-1\}} \frac{f(j)}{f_j}. 
	\]
\end{theorem}

See also \cite{bac06,barnog02,barnog06,cohire18,doukimozksoksol17}. 

\bigskip

{\bf Acknowledgements.} A significant part of the work of the first author on this chapter was done during his stay (August-December 2018) as a
visiting professor at Department of Mathematical Sciences at Purdue University Fort Wayne. His research was supported, in part, by a 
Bulgarian NSF contract DN02/2-2016.

\end{document}